\documentstyle[epsfig]{aipproc}

\begin{document}
\title{On the Role of External Constraints in a Spatially Extended
Evolutionary Prisoner's Dilemma Game}

\author{Gy\"orgy Szab\'o$^*$, Tibor Antal$^{\dagger}$,
P\'eter Szab\'o$^{\ddagger}$, and Michel Droz$^{\dagger}$}
\address{
$^*$Research Institute for Technical Physics and Materials Science \\
P.O.Box 49, H-1525 Budapest, Hungary \\
$^{\dagger}$Department of Theoretical Physics, University of Geneva,
1211 Geneva 4, Switzerland \\
$^{\ddagger}$Department of Ecology, J\'ozsef Attila University,
H-6721 Szeged, Egyetem u. 2, Hungary}

\maketitle

\begin{abstract}
We study the emergency of mutual cooperation in evolutionary
prisoner's dilemma games when the players are located on a square
lattice. The players can choose one of the three strategies:
cooperation ($C$), defection ($D$) or  ``tit for tat'' ($T$),
and their total payoffs come from games with the nearest neighbors.
During the random sequential updates the players adopt one of their
neighboring strategies if the chosen neighbor has higher payoff.
We compare the effect of two types of external constraints added to
the Darwinian evolutionary processes. 
In both cases the strategy of a randomly chosen player
is replaced with probability $P$ by another strategy.
In the first case, the strategy is replaced by a randomly chosen 
one among the two others, while in the second case the new strategy 
is always $C$. Using generalized 
mean-field approximations and Monte Carlo simulations the strategy
concentrations are evaluated in the stationary state for different
strength of external constraints characterized by the probability $P$.
\end{abstract}

\section*{Introduction}

The successful applications of game theory in the area of economics
and political decisions initiated its increasing development after the
second world war \cite{vNM}. Originally, the game theory is devoted to
find the optimal strategy for a given game between two intelligent
players. The straightforward developments involve the generalization
toward the iterated games of $n$ players with assuming local interactions
among the spatially distributed players. The spatial evolutionary
prisoner's dilemma games (SEPDG) has attracted a particular
attention because of its applicability in the human and behavior sciences
as well as in biology \cite{HS,sigmund,msmith,axelrod,weibull}.
Nowadays the prisoner's dilemma game is considered to be the metaphor for
studying the emergence of cooperation among selfish individuals.
The emerging cooperation appears to be crucial at many transitions
in evolution \cite{MSSz}.
The first numerical investigations have shown that the cooperation
can be maintained by very simple strategies in the iterated games
\cite{axelrod}. Very recently it is demonstrated that
the players can be as simple as bacteriophages (viruses that infect
bacteria)\cite{TC,NS99}.

In these systems the players wish to maximize their individual income
coming from games with other players. The prisoner's dilemma game is a
simple version of the two-player matrix games where the players' income
depend on their simultaneous choice between two options. Following the
widely accepted expressions each player can choose defection or cooperation
with the other player. The defector reaches the highest payoff $t$ (called
temptation to defect) against the cooperator, which receives then the lowest
reward $s$ (called sucker's payoff). For mutual cooperation [defection]
each player receives the same payoff $r$ (reward for mutual cooperation)
[$p$ (punishment)]. The game is
symmetric in the sense that player's income is independent of the
player itself, it depends only on their choice.  The mentioned payoff
values satisfy the inequalities $t > r > p > s$ and
$2r > t+s$. These assumptions provide the largest total payoff for the
mutual cooperators. Comparing to this situation the defector reaches
extra income against the cooperator whose loss exceeds the defector's
benefit. Consequently, the choice of defection can be interpreted as
an exploiting behavior. These are the main features for which the prisoner's
dilemma games are used to study the emergence of mutual cooperation,
altruism and ethic norms among selfish individuals \cite{axelrod,NMS}.

The rational players should defect as this choice provides the
larger income, independently of the partner's decision.
However, this situation creates a dilemma for intelligent players 
as mutual cooperation would result in higher income
for each of them than mutual defection does.

In the iterated round-robin prisoner's dilemma games we can introduce
some simple evolutionary processes without assuming intelligent
players (who are capable to find the best strategy if it exists).
These games are started from an initial set of strategies, which
defines the player's decision in the knowledge of their previous
choices. The evolutionary process is devoted to model the Darwinian
selection principle among $n$ (selfish) players whose total
income comes from $n-1$ games within a given round. In the simplest
evolutionary models the worst player adopts the winner's
strategy round by round.

The numerical simulations have demonstrated
the crucial role of the so-called ``tit for tat'' strategy
in the emergence of mutual cooperation. Despite of its simplicity
the ``tit for tat'' strategy won the computer tournaments conducted
by Axelrod \cite{axelrod}. The ``tit for tat'' strategy cooperates
in the first step and then always repeats his co-player's previous
decision. This strategy cooperates forever with all the other
so-called nice strategies which never defect first. Furthermore,
its defection and cooperation can be interpreted as a punishment
and forgiveness when reacting to the previous decision for other
strategies. The most remarkable feature of this strategy is that
it is capable to sustain the mutual cooperation among themselves 
in the presence of defectors.

Early numerical investigations have also indicated the
importance of local interactions because it favors the
formation of cooperating colonies. In the simplest models
the players are distributed on a lattice and the interaction
(the games between two player as well as the strategy adoption)
is limited to a given neighborhood. Evidently, the short range
interactions enhance the role of fluctuations at the same time.
These models can be well investigated by sophisticated
methods of non-equilibrium statistical physics. 

For the numerical investigation of the spatial effects Nowak
and May~\cite{NM} have introduced an SEPDG model, which is
equivalent to a two-state cellular automaton. 
Each lattice site can be in one of the two states $C$ and $D$,
representing the two simple strategy ``always cooperate'' and 
``always defect'' respectively.
The income for a given player (site) comes from games with its neighbors
(and also with itself in some version of the model). 
According to the cellular automaton rule the
players modify their strategy simultaneously in discrete time steps.
Namely, each player adopts the best strategy found in its neighborhood.
The step by step visualization of the strategy distribution on a
two-dimensional lattice exhibits different spatio-temporal patterns
(homogeneous and coexisting strategies, transitions between these states,
competing interfacial invasions, etc.) depending on the payoff matrix,
which is characterized by a single parameter. In these models the randomness
is restricted to the initial states. In a subsequent work  Nowak
{\it et al.} \cite{NBM} have extended the former models by allowing
irrational strategy adoptions with some probability. The simulations
indicated that the randomness favors the spreading of $D$ strategies.
These results have initiated systematic numerical investigations
of many stochastic cellular automata \cite{KD96,KD98,KDK,CO}.

The study of spatio-temporal patterns observed in nature, however,
requires continuous time description \cite{HG,NM}. Moreover, it is
difficult to analyze the above mentioned stochastic cellular automata 
in the framework of generalized mean-field approximation,
which is often used in non-equilibrium physics.
To reduce the technical difficulties Szab\'o and T\H oke have suggested
a simplified dynamics \cite{epdg2s}. The systematic investigations
of this model have justified that when tuning the model parameters
the stationary state undergoes two consecutive phase transitions
which belong to the directed percolation (DP) universality 
class\cite{epdg2s,CO}. Very recently this SEPDG model has been
extended by allowing three strategies for the players \cite{epdg3s}.
In the present work this three-strategy model will be compared with
its simplified version. During the model descriptions and discussion,
our attention will be focused on the motivations, the elementary processes
and their consequences as well as on the universal features relating the
SEPDGs to the area of complex systems.

\section*{Spatial evolutionary model with three strategies}

In the present spatial evolutionary prisoner's dilemma game the
players are located on the sites ${\bf x}=(i,j)$ of a square lattice,
where $i,j=1, \ldots , L$. To avoid the undesired boundary effects we
assume periodic boundary conditions. 
Each player follows one of the three strategies: $D$ defects always;
$C$ cooperates unconditionally; $T$ accommodating to the partner's strategy
chooses defection against $D$ and cooperation with $C$ and $T$. In fact
the name $T$ refers to the strategy ``tit for tat'' which first cooperates
and later repeats the partner's previous decision. Consequently,
after the first step the decisions of these two strategies are equivalent
against $C$, $D$ and themselves.
The consequences of the different first decisions become irrelevant if the
strategy changes (defined below) are rare comparing to the frequency of
games. At the site ${\bf x}$ the player's
strategy is denoted by a three-component unit vector whose possible
values are
\begin{equation}
{\bf s}({\bf x})=\left( \matrix{1 \cr 0 \cr 0 \cr}\right)~,~~~
    \left( \matrix{0 \cr 1 \cr 0 \cr}\right)~,~~~
    \left( \matrix{0 \cr 0 \cr 1 \cr}\right)~
\end{equation}
corresponding to the $D$, $C$, and $T$ strategies respectively.
At a given time the state of the whole system is described by
the variables ${\bf s}({\bf x})$.

For each player the total payoff comes from the games with its four
nearest neighbors. Using the above formalism the total payoff
$m({\bf x})$ for the player at site ${\bf x}$ is given as
\begin{equation}
m({\bf x})=\sum_{\delta {\bf x}} {\bf s}^{\star}({\bf x}){\bf M}{\bf s}
({\bf x}+\delta{\bf x})
\end{equation}
where ${\bf s}^{\star}({\bf x})$ is the transpose matrix of 
${\bf s}({\bf x})$ and
the summation runs over the four nearest neighbors ($\delta 
{\bf x}$).  Accepting the simplified payoff matrix suggested
by Nowak and May \cite{NM} ${\bf M}$ is given by the following
expression:
\begin{equation}
{\bf M}=\left( \matrix{0 & b & 0 \cr
                       0 & 1 & 1 \cr
                       0 & 1 & 1 \cr}\right)
\end{equation}
where the only free parameter $b$ ($1<b<2$) measures the temptation
to defect. In the above mentioned notation the present payoff matrix
corresponds to the choices: $r=1$, $p=0$, $t=b$, and $s=-\varepsilon$
in the limit $\varepsilon \to 0$.

To model the Darwinian selection rule the players are allowed to modify
their strategy. In the simplest case the system evolution is governed
by random sequential updates. It means that a randomly chosen player
(e.g. at site ${\bf x}$) adopts one of its neighboring strategy,
${\bf s}({\bf x}+\delta {\bf x})$, if $m({\bf x}+\delta {\bf x}) 
> m({\bf x})$ and this elementary process is iterated many times.

\begin{figure}[h!] 
\centerline{\epsfig{file=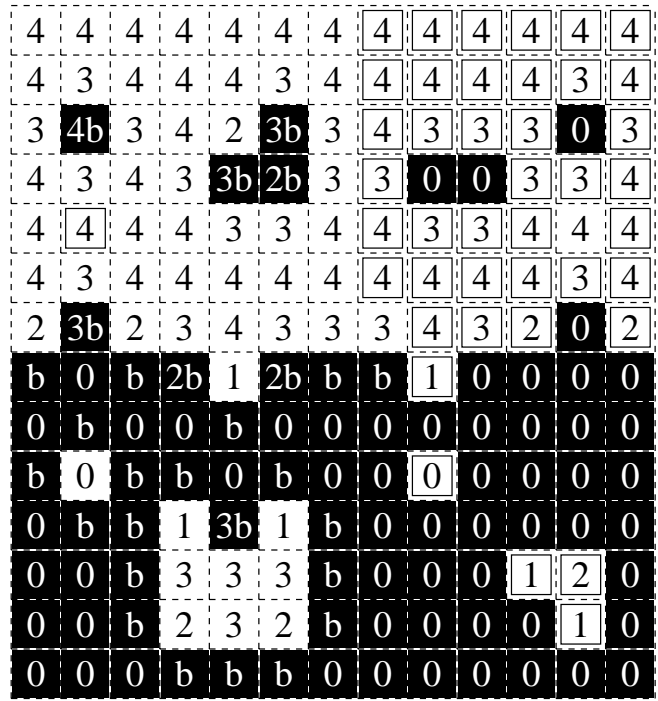,width=3.5in}}
\vspace{10pt}
\caption{The payoff distribution is indicated by figures on the square
lattice for a given configuration of $D$ (black box), $C$ (empty), 
and $T$ (open box) strategies.}
\label{fig:podistrn}
\end{figure}

Here it is worth mentioning that a state consisting only of $C$ and
$T$ strategies leads to a uniform payoff distribution [$m({\bf x})=4$]
and the above dynamics leaves this state unchanged. 
An example of a more complicated situation is given in Figure 
\ref{fig:podistrn}. The payoffs associated with the three different 
strategies, $D$, $C$ and $T$, are explicitly given.

The reader can easily check that inside a $D$ region the defectors receive
zero payoffs. The same is true for a solitary $T$ surrounded by defectors. 
In the absence of $C$ strategies, however, two (or more) neighboring $T$
strategies will invade the $D$ territories because their mutual cooperation
gives them some incomes, while the defectors' payoff remain zero. 

\begin{figure}[h!] 
\centerline{\epsfig{file=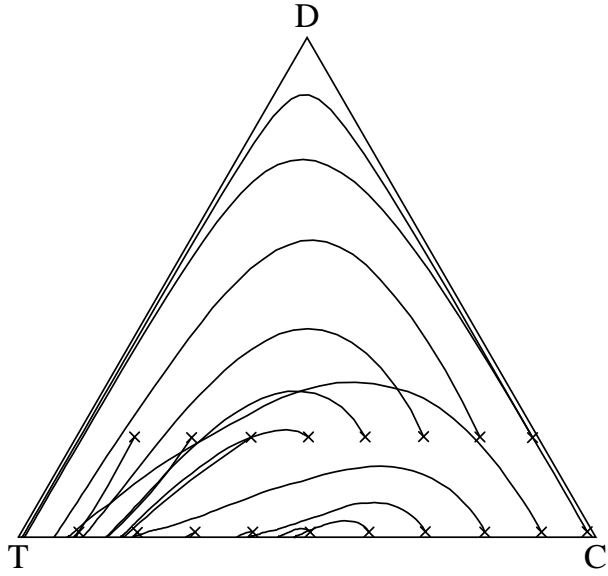,width=3.5in}}
\vspace{10pt}
\caption{Monte Carlo results for the time evolutions in the absence of
mutation if the system is started from different (uncorrelated) initial
states indicated by X symbols.}
\label{fig:flowp}
\end{figure}

In the presence of $C$ strategies, however, the above situation becomes
quite different as the exploitation provides large incomes for the 
defectors. As a result, the defectors can invade the neighboring $C$
or $T$ sites for some configurations. This process dominates the time
evolution for small $T$ and large $C$ concentrations as illustrated in
a ternary diagram (see Figure~\ref{fig:flowp}). 
Note that the trajectories are two dimensional projections of
a many dimensional space. Accordingly, there can be crossing of 
trajectories. As the average defector's
payoff decreases with the $C$ concentration, sooner or later
the $T \to D$ invasion processes will govern the system evolution and,
finally, all the $D$ strategies extinct. Figure~\ref{fig:flowp} shows
clearly that the ratio of $C$ and $T$ strategies in the final (frozen) 
state depends on the initial state.

It is emphasized that in the absence of $T$ strategies the defectors will
dominate the present system in the final state. It is not evident as in
Figure~\ref{fig:podistrn} one can find many $D$-$C$ pairs where $C$ beats
$D$. In general, these pairs are located along the horizontal and vertical
straight fronts separating the $D$ and $C$ domains. The random sequential
invasions, however, makes the smooth fronts irregular and this situation
generally prefers the $D \to C$ invasion to the opposite one. As a result, the 
``sharp'' $D$ fronts cut the $C$'s domains into small pieces and finally all
the $C$s will be eliminated.

The reader can easily recognize that in most of the $C$-$D$ (or $T$-$D$)
confrontations the direction of dominance is not affected by the value of
$b$ within the prescribed region ($1<b<2$). The systematic analysis shows
that there is only one situation when the value of $b$ becomes important.
Namely, if a defector has a payoff of $2b$ while its $C$ (or $T$) 
counterpart has 3. In this case, $D$ wins if $b>3/2$, otherwise $D$ will
be invaded. These types of elementary processes, however, do not modify
the system behavior drastically \cite{epdg3s}, therefore the subsequent
investigations will be focused on the case $b>3/2$.

The above dynamical rules introduce some noises (irrational choices)
in the system evolution. Now an additional (superimposing) noisy term is
introduced by allowing the appearance of mutants with probability $P$.
In fact, the effect of two different external constraints 
(mutation mechanisms) will be studied in models A and B as a function of $P$.

\subsection*{Model A}

In the first model, the above evolutionary rule is modified as follows.
Each randomly chosen player adopts with probability $P$ a
randomly chosen strategy among the two other strategies.
With probability $1-P$ it follows the old rule.

This model can describe the behavior of those biological and economical
problems where the appearance of mutants cannot be neglected \cite{msmith,HS}.
The main feature of this model is that this mutation mechanism does not
allow the extinction of any strategy.

\subsection*{Model B}

In the second model the mutation mechanism is restricted to the adoption
of $C$ strategies \cite{epdg3s}. In other words, the randomly chosen player
adopts the $C$ strategy with probability $P$, otherwise it adopts one of
its neighboring strategy if this neighbor has higher income. 
Note that in this case
the extinction of the $D$ and/or $T$ strategies is permitted.

Model B is devoted to describe the effect of an external constraint
which enforces the cooperative behavior naively by supporting an
unconditional cooperation. Such a phenomenon can be observed in
human societies in which any kind of social pressure enforces 
the $D$ and $T$ players to choose the $C$ strategy.
Furthermore, a $T$ player surrounded by only cooperating strategies
($C$ or $T$) is motivated  to adopt the $C$ strategy also because of its 
convenience. In fact, playing $C$ is simpler than playing $T$,
which requires the knowledge of the previous decision of your neighbors.

\section*{Mean-field approximation}

In the classical mean-field approximation the system is described
by the strategy concentrations which satisfy the normalization
condition $c_D(t)+c_C(t)+c_T(t)=1$. In this approach the average
payoffs are given as:
\begin{eqnarray}
m_D&=&b c_C \ ,\nonumber \\ 
m_C&=&c_C+c_T \ 
\label{eq:mfpayoff} ,\\
m_T&=&c_C+c_T \ . \nonumber
\end{eqnarray}
For model A, the time dependent concentrations satisfy
the following equations of motion:
\begin{eqnarray}
\dot{c}_D&=&{P \over 2}(c_C+c_T-2c_D) \mp (1-P) c_D(c_C+c_T) \ , \nonumber \\
\dot{c}_C&=&{P \over 2}(c_T+c_D-2c_C) \pm (1-P) c_D c_C \ ,
\label{eq:mfa} \\
\dot{c}_T&=&{P \over 2}(c_D+c_C-2c_T) \pm (1-P) c_D c_T \ , \nonumber
\end{eqnarray}
where the upper (lower) signs are valid if $m_D < m_C=m_T$
($m_D > m_C=m_T$). In these expressions the first terms describe
the effect of external constraint, the second terms come from the Darwinian
selection mechanism.

After some algebraic manipulations one can easily get the following
stationary solution (for $P<1$):
\begin{eqnarray}
c_D&=&{1+P/2-\sqrt{1-P+9P^2/4} \over 2(1-P)} \ ,\nonumber \\ 
c_C&=&c_T={1-c_D \over 2} \ \ .
\label{eq:MFsolA}
\end{eqnarray}
Here all the three strategies are present for arbitrary values of $P$.
Notice that the concentrations of $C$ and $T$ strategies are the same
due to the symmetries of Eqs.~(\ref{eq:mfa}).
In the limit $P \to 0$, however, the concentration of $D$ strategy vanishes. 
Evidently, the concentration of the three strategies becomes equal
when the evolution is governed exclusively by the mutation ($P=1$).

For model B the corresponding equations of motion are similar to those
given by Eqs.~(\ref{eq:mfa}), the differences appear in the first terms
proportional to $P$. Namely,
\begin{eqnarray}
\dot{c}_D&=&-Pc_D \mp (1-P) c_D(c_C+c_T) \ , \nonumber \\
\dot{c}_C&=&+P(c_T+c_D) \pm (1-P) c_D c_C \ ,
\label{eq:mfb} \\
\dot{c}_T&=&-P2c_T \pm (1-P) c_D c_T \ , \nonumber
\end{eqnarray}
where the average payoff values are given by Eqs.~(\ref{eq:mfpayoff})
and the conditions of validity of the upper and lower signs are defined as above.   
The analytical solution of these equations predicts strikingly different
behavior in the stationary state \cite{epdg3s}, that is, for $0<P<1/2$ 
\begin{eqnarray}
c_D&=&{1-2P \over 1-P} \ ,\nonumber  \\ 
c_C&=&{P \over 1-P} 
\label{eq:MFsolB}\ ,  \\
c_T&=&0 \ ,\nonumber
\end{eqnarray}
while the system goes to the absorbing state ($c_C=1$ and
$c_D=c_T=0$) for $P>1/2$. The most surprising result is the extinction of
$T$ strategy if $P > 0$.

\begin{figure}[h!] 
\centerline{\epsfig{file=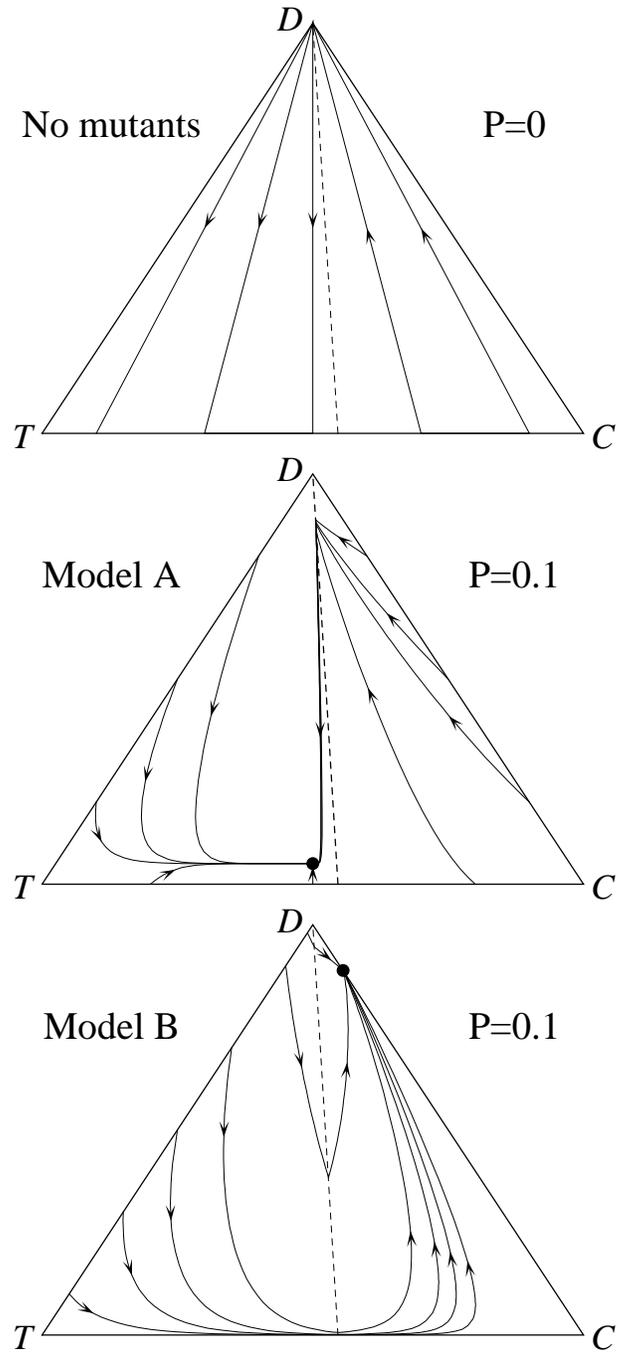,width=3.5in}}
\vspace{10pt}
\caption{Trajectories describing the time evolution in ternary diagrams
for respectively the model with no mutation, model A ($P=0.1$) and
model B ($P=0.1$). The dashed lines separate the regions where $D$
dominates or is dominated. The fixed points are denoted by bullets.}
\label{fig:flowzab}
\end{figure}

We have to emphasize the non-analytical behavior in the limit $P \to 0$.
As illustrated in the upper plot of Figure~\ref{fig:flowzab}, without the
mutation ($P=0$) the system evolves toward either a homogeneous $D$
state ($c_D=1$) or a mixed state composed of $C$ and $T$ strategies
with a ratio depending on the initial conditions. However, the homogeneous
$D$ state is unstable against $T$ invasions, therefore in the close vicinity
of this state some small perturbations can drive the system toward the state
of $C$+$T$. Conversely, this mixed state becomes unstable at a given
concentration (where $m_D=m_C=m_T$) against small perturbations increasing
$c_C$ and $c_D$ simultaneously. In other words, the system evolves toward
the $D$ dominance when the state is positioned on the right hand side of
dashed line (see the upper plot in Figure~\ref{fig:flowzab} as a results
of fluctuations. This feature explains why the system is so sensitive
to applied external constraints.
 
Figure~\ref{fig:flowzab} illustrates that for model A the mutation
drives the (concentration) trajectories away from the boundaries. On the right
hand side of the dashed line ($m_D>m_C=m_T$), $c_C$ and $m_D$ decrease
while $c_D$ and $c_T$ increase until one crosses the dashed line.
On the left hand side all the initial states tend toward the only fixed
point given by Eq.~(\ref{eq:MFsolA}).
For model B, however, there is no fixed point on the left hand side.
In this region the external insertion of $C$ strategies increases the
value of $c_C$ until $m_D$ becomes larger than $m_C=m_T$ and then the
$D$ invasion drives the system toward the fixed point defined by
Eq.~(\ref{eq:MFsolB}). During the $D$ invasion the external constraint
can compensate only the loss of $C$ strategies.
Consequently, the $T$ strategies die out exponentially fast.

Notice that the variation of $b$ leaves the fixed points unchanged, 
but modifies only the slope of the dashed line separating the two regions
mentioned above.

Within the framework of mean-field theory, the extinction of $T$
strategies is a consequence of the fact that here $m_T=m_C$
[see Eq.~(\ref{eq:mfpayoff})] in contrary to the 
spatially extended case, as illustrated in Figure~\ref{fig:podistrn}.

\section*{Monte Carlo simulations}

Systematic Monte Carlo simulations have been performed on a square lattice
consisting of $L \times L$ sites with periodic boundary conditions,
$L$ varying from 200 to 1500. The larger sizes were
used in the vicinity of the critical points. Each run started
from a random initial state. During the simulations we have monitored
the number of players playing a given strategy 
($N_{\alpha}$; $\alpha = D$, $C$ or $T$) and the payoffs related to a 
given strategy. After some relaxation time we have determined the average
concentrations
\begin{equation}
c_{\alpha} = \langle N_{\alpha}\rangle / L^2
\label{eq:c}
\end{equation}
and the fluctuations
\begin{equation}
\chi_{\alpha}=L^2 \langle (N_{\alpha}/L^2 - c_{\alpha})^2 \rangle
\label{eq:chi}
\end{equation}
by averaging over a sampling time varying between $10^4$
and $10^6$ Monte Carlo steps (MCS) per sites. The results obtained 
respectively for model A and B are the following.

\subsection*{Results for Model A}

\begin{figure}[h!] 
\centerline{\epsfig{file=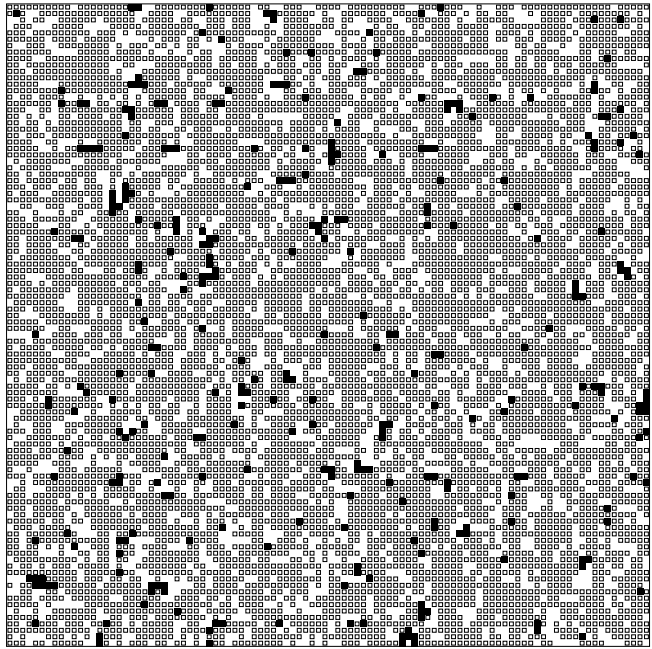,width=3.5in}}
\vspace{10pt}
\caption{Typical stationary state distribution of defector 
(closed black square), cooperator
(white area), and Tit for Tat (open square) strategies in model A,
on a $100 \times 100$ portion of a large system for $P=0.02$ and $b>3/2$.}
\label{fig:assp02}
\end{figure}

Figure \ref{fig:assp02} shows a typical strategy distribution for
the stationary state at a small value of $P$. In contrary to the
mean-field prediction [see Eqs.~(\ref{eq:MFsolA})] the
system is dominated by the $T$ strategies. The randomly inserted 
$D$ and $C$ strategies form small islands. Occasionally the larger
$C$ islands are occupied by $D$s, however, a consecutive $T$ invasion
will eliminate the larger $D$ territories and maintains the $T$ dominance.
At the same time this process prevents the formation of large $C$ islands
inside a $T$ domain.

\begin{figure}[h!] 
\centerline{\epsfig{file=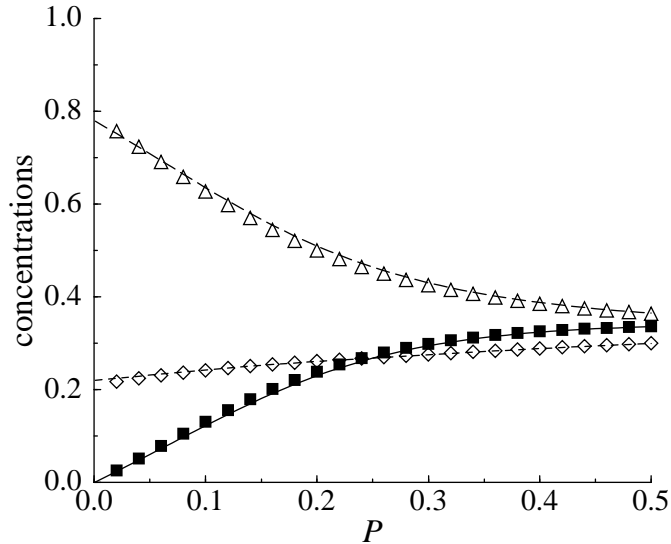,width=3.5in}}
\vspace{10pt}
\caption{Stationary strategy concentrations in model A 
as a function of $P$ for $b>3/2$. The Monte Carlo
results are represented by closed squares ($D$),  open diamonds ($C$),
and open triangles ($T$). The prediction of pair approximation is indicated
by solid, dashed, and long-dashed lines respectively.}
\label{fig:s3mmc}
\end{figure}

One can observe in Figure~\ref{fig:s3mmc} that when increasing the value 
of $P$, the concentration of $D$ and $C$
strategies increases monotonously. In the limit $P \to 1$, the 
strategy distribution on the lattice tends toward a random (uncorrelated)
one $c_D=c_C=c_T=1/3$ in agreement with the classical mean-field theory
[see Eq. (\ref{eq:MFsolA})]. In this case, instead of the neighbor invasions,
the system evolution is ruled by the stochastic mutation mechanism.

As shown in Figure \ref{fig:s3mmc}, the Monte Carlo data agree remarkably well
with the results of the pair approximation. This pair approximation is
considered as a generalized mean-field theory taking the nearest-neighbor
correlations explicitly into account. The details of this calculation
are available in many previous works \cite{epdg2s,epdg3s,ST}.
The good agreement refers to the absence of long-range correlations which
is observable in the ``homogeneous'' strategy distribution (see
Figure~\ref{fig:assp02}). It is worth mentioning that the pair approximation
is capable to describe the dominance of $T$ strategies in the limit
$P \to 0$.

\subsection*{Results for Model B}

\begin{figure}[h!] 
\centerline{\epsfig{file=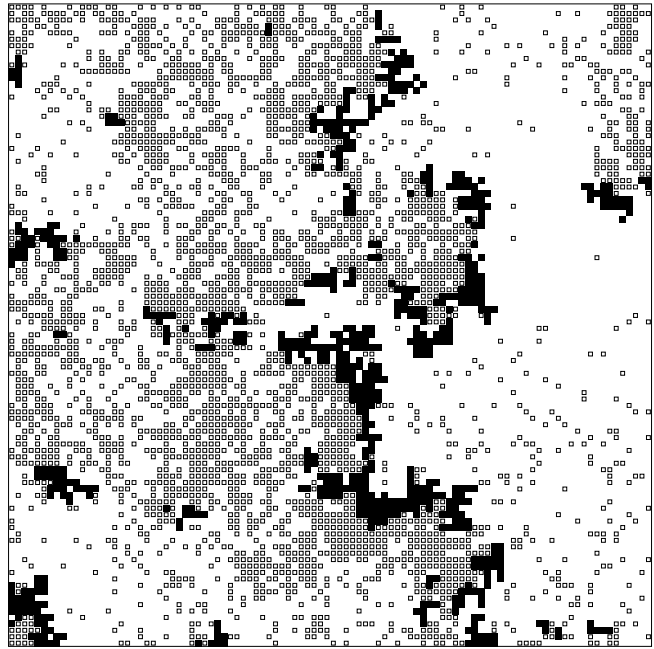,width=3.5in}}
\vspace{10pt}
\caption{Typical snapshot for the stationary state of model B at $P=0.02$
and for $b>3/2$. The symbols are the same as in Figure~\ref{fig:assp02}.}
\label{fig:bssp02}
\end{figure}

In order to visualize the relevant differences between the two models
at small $P$ values the strategy distribution for model B is displayed
in Figure~\ref{fig:bssp02}. When comparing the corresponding snapshots 
(Figures~\ref{fig:assp02} and \ref{fig:bssp02}) the reader can easily
recognize the most striking differences. Namely, the appearance of a
strongly correlated spatial structure for model B. In this case the
formation of large $C$ domains inside the sea of $T$ strategies is not
prevented by the random appearance of $D$ mutants as happened in the 
previous case. The large $C$ domains (white areas in
Figure~\ref{fig:bssp02}), however, are unprotected against the $D$
invasion. Figure~\ref{fig:bssp02} shows some $D$ domains (black areas)
invading the $C$'s territories. These $D$ domains are ``strip-like''
because their territories are invaded simultaneously by the $T$ strategies.
This invasion process is similar also for larger values of $P$
(but $0<P<P_{c1}$, see later), and only the average invasion velocity
changes. On the other hand, the randomly
inserted $C$ strategies survive
and accumulate in the $T$ domains. Consequently, far behind the $T$-$D$
invasion front the $T$'s territory will be occupied by the externally
inserted $C$s and then this area becomes unprotected against the $D$
invasion. Sooner or later this area will be invaded by $D$s and the
above process repeats itself. This means that the cyclic invasion
maintains a self-organizing domain structure. Here we have to emphasize
that this cyclic (rock-scissors-paper game like) dominance is provided 
by this external constraint.

Similar processes are observed in the forest-fire models \cite{GK,DS}
introduced by Bak et al. \cite{BCT} to model the phenomenon of
self-organized criticality. In these models each cell can be in one of the 
following three states: non-burning tree, burning tree and ash.
The dynamics are governed by cyclic dominance, similarly to 
our model B. Note that the consequence
of cyclic invasion with three (or more) states are studied in Lotka-Volterra
models \cite{Lotka,Volterra,AD,ST} and in cyclically dominated voter models
\cite{TI,Tain93,T94,FKB,SSM}.

In model B the transition from the $T$ to $C$ state introduces a characteristic
length and time unit, both proportional to $1/P$. In other words, this
length unit is characteristic to the typical (linear) size of the $T$+$C$
domain, and the time unit corresponds to periodic time of cyclic invasion
processes at a given site.

\begin{figure}[h!] 
\centerline{\epsfig{file=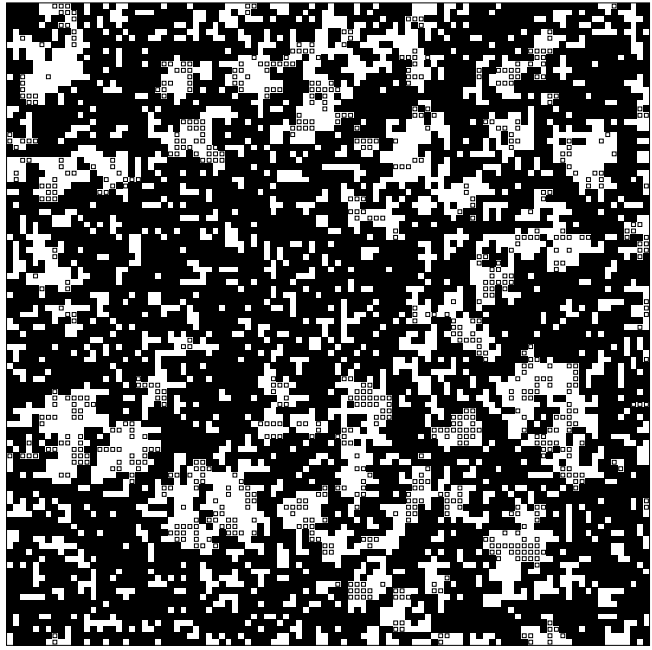,width=3.5in}}
\vspace{10pt}
\caption{Stationary strategy distribution in model B for $P=0.13$
and $b>3/2$. The symbols are the same as in Figure~\ref{fig:assp02}.}
\label{fig:bssp13}
\end{figure}

When increasing the value of $P$, the typical size of $T$+$C$ domains
decreases and the concentration of $D$'s increases. It is found that
the $T$ strategies die out if $P>P_{c1}=0.1329(1)$. Figure~\ref{fig:bssp13}
shows a typical snapshot in the vicinity of this critical value. In this
case the external support is sufficiently strong to maintain small $C$
clusters inside the $D$ domains. The most remarkable feature of this
snapshot is that the $T$'s form non-uniformly distributed small
(isolated) colonies. The observation of time evolution of configuration
shows that these $T$ colonies walk randomly, they can extinct spontaneously,
a single colony can split into two, or two colonies can merge. This
phenomenon is analogous to the branching annihilating random walks (BARW)
exhibiting a critical transition when varying the control
parameters \cite{CT96,CT98}. The corresponding critical transitions,
both for our model B and for BARW,
belongs to the so-called directed percolation universality
class \cite{Janssen,Grassb}.

\begin{figure}[h!] 
\centerline{\epsfig{file=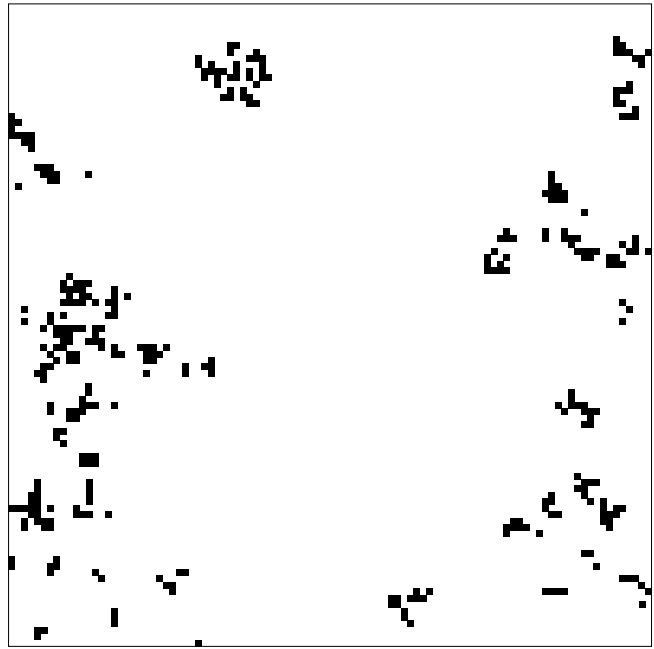,width=3.5in}}
\vspace{10pt}
\caption{Stationary distribution of $C$ (white area) and $D$ (black boxes) 
strategies in model B for $P=0.366$ and $b>3/2$.}
\label{fig:bssp366}
\end{figure}

For $P > P_{c1}$, the concentration of $D$ decreases monotonously if $P$
is increased and vanishes at $P=P_{c2}=0.3678(1)$. This extinction
process is similar to the previous one, i.e.\ it also belongs to
the DP universality class. The similarity in the correlations is
recognizable in the spatial distribution of the extincting
strategies when comparing the snapshots displayed in
Figures~\ref{fig:bssp13} and \ref{fig:bssp366}.

For $P > P_{c2}$, any initial state evolves toward the
absorbing state where all the players follow the $C$ strategy.

\begin{figure}[h!] 
\centerline{\epsfig{file=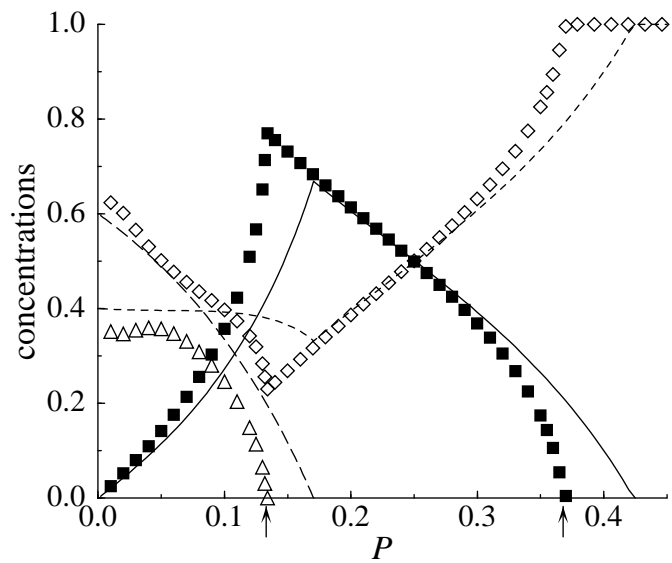,width=3.5in}}
\vspace{10pt}
\caption{Simulation and pair-approximation results for the stationary
concentration of strategies in model B, versus $P$.
The notation agrees with those of Fig. \ref{fig:s3mmc} and the arrows
indicate the critical points where $T$ and $D$ strategies extinct.}
\label{fig:s3mc}
\end{figure}

The results of our systematic investigations are summarized in
Figure~\ref{fig:s3mc}. Systematic numerical investigations
in the close vicinity of the critical points show that the vanishing
concentrations follow the same power law behavior. Namely,
\begin{eqnarray}
c_T&=&(P_{c1}-P)^{\beta} \ , \nonumber  \\ 
c_D&=&(P_{c2}-P)^{\beta} \ ,
\label{eq:beta}
\end{eqnarray}
in the limits $P_{c1}-P \to 0$ and $P_{c2}-P \to 0$ respectively
and $\beta = 0.57(3)$ in both cases \cite{epdg3s}. Within the
statistical error this value of the exponent $\beta$ agrees with
the one of the 2+1 dimensional directed percolation \cite{BFM,JFD}.

As expected, these critical transitions are accompanied with the
divergence of concentration fluctuations, i.e.
\begin{eqnarray}
\chi_T&=&(P_{c1}-P)^{-\gamma} \ , \nonumber  \\ 
\chi_D&=&(P_{c2}-P)^{-\gamma} \ ,
\label{eq:gamma}
\end{eqnarray}
in the vicinity of the corresponding critical points. The numerical
fitting yields $\gamma =0.37(9)$ in agreement with the 
DP values \cite{BFM,JFD,Janssen,Grassb}.

Despite the same universal behavior there is a remarkable difference
between the two extinction processes. The second extinction process
(at $P=P_{c2}$) results in a frozen (time independent) absorbing state.
Conversely, the transition at $P=P_{c1}$ is an example where the
extinction of $T$ strategies happens on a fluctuating background.
In other words, the properties of the absorbing state (frozen or fluctuating)
do not affect the critical behavior of our model.

As demonstrated in Figure~\ref{fig:s3mc} the results of Monte
Carlo simulations are reproduced qualitatively well by the
pair approximation \cite{epdg3s}. The striking differences
are related to the long-range correlations accompanying the
critical transitions at $P=P_{c1}$ and $P_{c2}$. Due to the strongly
correlated domain structure, illustrated in Figure~\ref{fig:bssp02},
the largest deviation can be observed for small $P$ values. 
We note that the concentration fluctuations, defined by
Eq.~(\ref{eq:chi}), also diverge in the limit $P \to 0$. Unfortunately,
in this particular case,
we could not deduce a reliable value for the exponent $\gamma$
because of the significant size effects. Further systematic
analyzes are required to clarify what happens in this limit.

\section*{Conclusions}

We have studied quantitatively the effect of external constraints on the
emergence of cooperation in an evolutionary prisoner's dilemma game with
three possible strategies (cooperation, defection and tit for tat). In the
present spatial model the players are distributed on a square lattice
and their interactions are restricted to nearest neighbors. The Darwinian
selection rule is modeled by the adoption of the neighboring successful
strategies. This evolutionary process is superimposed by two types of
mutation mechanisms (external constraints) whose strength is
characterized by a control parameter $P$.

The choice of these three possible strategies yields non-analytical 
behavior in the limit $P \to 0$ for both the mean-field approximation
and Monte Carlo simulation. The time-dependent predictions of mean-field
theory are sensitive to the small perturbations.

According to the Monte Carlo simulations, in the absence of external constraint
the system tends toward a frozen state composed from $C$ and $T$ strategies
whose ratio depends on the initial concentrations. For both types of
external constraints (models A and B) the system evolves toward a
stationary state independently of the initial condition, and the
defector concentration vanishes linearly as $P\to 0$. In the limit
$P \to 0$, however, model A and B will exhibit different ratio
of $C$ and $T$ strategies. This difference is related to the appearance
of self-organizing patterns for model B. The present investigation
indicates that such a society of strategies (or species) are very
sensitive to the type of external supports (or the ration of mutation
rates).

\begin{figure}[h!] 
\centerline{\epsfig{file=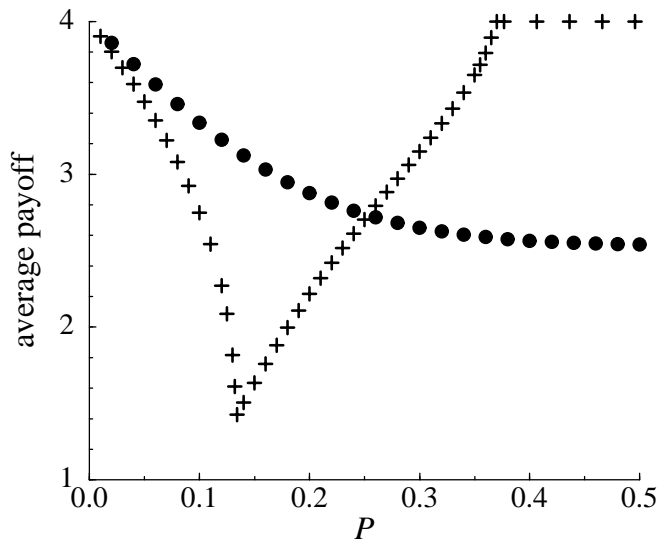,width=3.5in}}
\vspace{10pt}
\caption{Comparison of average payoffs as a function of $P$ for the
models A (closed circles) and B (plus symbols). The sharp minimum 
coincides with the extinction of $T$ strategies for model B.}
\label{fig:apayoff}
\end{figure}

The measure of mutual cooperation can be well characterized by the average
payoff whose maximum (4) can be reached only in the absence of defectors.
Figure~\ref{fig:apayoff} compares the Monte Carlo results for the models
A and B. Surprisingly, for weak external support (small $P$) the average
payoff is larger for model A than for model B. In contrary to the
naive expectation, the weak support of defenseless cooperators results in
opposite consequence. Namely, this mechanism feeds the defectors and
simultaneously prevents their elimination by the retaliatory ($T$)
strategies. 

Examples from the political and economical world justifies the above
conclusions. In general, the exploiters are preferred by the governmental
support for the defenseless layer of a society. The most dangerous effect
is the reduction in the $T$ type population which can maintain the
mutual cooperation against the exploiters. From the view point of
cooperation, it is better to help those individuals who are able to
prevent themselves against the exploitation.

Evidently, for sufficiently large $P$ values the random insertion 
of $C$ strategies can provide their dominance. In this case the A type
external support is preferred to the B one if we wish to improve
cooperation. Above a threshold value this type of external constraints 
yields a homogeneous $C$ state which is defenseless against any defector
appearing occasionally in a real system. Further systematic research
is required to clarify what happens in those models where the mutation
mechanism is characterized by three independent control parameters.

The present study confirms that the $T$ strategy is able to prevent
the spreading of defection in the spatial models. We have to emphasize, 
however, that according to the simplest mean-field theory, $T$ dies
out if the external support is of B type.
Consequently, the defectors will dominate those systems where the
mean-field theory is exact (e.g. infinite range of interaction, or
randomly chosen partnership). In these mean-field like systems, the games between
the ``parent`` and ``its offspring'' is not emphasized
(they are not neighbors), which is an advantage for the defectors
comparing to spatially extended models.
In the light of this feature our investigations
imply many interesting questions related to the transition from the
``short range'' spatially extended systems to the ``long range''
of mean-field like ones.

\vspace{3mm}
\paragraph*{Acknowledgments.}
This work was supported by the Hungarian National Research Fund
under Grant No. T-23552 and by the Swiss National Foundation.

\end{document}